\documentclass[twocolumn,amssymb,fleqn, twocolumns]{revtex4} 
\usepackage{epsfig,amssymb,amsmath,graphicx,subfigure,hyperref  }  
\usepackage{color}

\usepackage{multirow} 
\usepackage{tabularx}

\begin{document}

\title{Stress driven fractionalization of vacancies in regular packings of elastic particles}
\author{Zhenwei Yao}
\email{zyao@sjtu.edu.cn}
\affiliation{School of Physics and Astronomy, and Institute of Natural
Sciences, Shanghai Jiao Tong University, Shanghai 200240, China}
\begin{abstract}
  Elucidating the interplay of defect and stress at the microscopic level is a
  fundamental physical problem that has strong connection with materials
  science.  Here, based on the two-dimensional crystal model, we show
  that the instability mode of vacancies with varying size and morphology
  conforms to a common scenario. A vacancy under compression is fissioned into
  a pair of dislocations that glide and vanish at the boundary. This
  neat process is triggered by the local shear stress around the vacancy. The
  remarkable fractionalization of vacancies creates rich modes of interaction
  between vacancies and other topological defects, and provides a new dimension
  for mechanical engineering of defects in extensive crystalline structures. 
\end{abstract}

\maketitle

\section{Introduction}

Topological defects are emergent structures commonly seen in various ordered
condensed media~\cite{chaikin00a,pertsinidis2001diffusion,
Bowick2009c,koning2014crystals,yao2014polydispersity}.  As a fundamentally
important crystallographic defect, vacancies are highly involved in several
important physical processes in both two-~\cite{lee2005diffusion, meng2014elastic,yao2017topological} and three-dimensional
systems,~\cite{jhi2001vacancy, maiti2015free} such as in
facilitating migration of
atoms,~\cite{shackelford2005introduction,yao2014dynamics} and crystallization of
DNA-programmable nanoparticles,~\cite{knorowski2012dynamics,girard2019particle}
and in resolving geometric frustrations in curved
crystals~\cite{Bausch2003e,irvine2010pleats,meng2014elastic,yao2017topological,guerra2018freezing}.
Vacancies, together with other defects, are crucial for the characteristic of
heterogeneous stress distributions in the packing of
particles,~\cite{knorowski2012dynamics, meng2014elastic, yao2017topological,
van2018revealing} which has connections with the formation of force chain
structures and the resulting jamming transition in granular
media~\cite{liu1998nonlinear, peters2005characterization, majmudar2005contact,
porter2015granular,laubie2017stress}.  Understanding the interplay of vacancy
and stress yields insights into a variety of crystallographic
phenomena~\cite{jaeger1992physics,geng2001footprints,aranson2006patterns,meng2014elastic}.
While these problems have been extensively studied from the statistical
perspective,~\cite{o2001force, majmudar2005contact,maiti2015free} the
microscopic behavior of vacancies in stressed particle arrays is still not
clear.  Furthermore, vacancies in particulate systems tend to suppress the
relaxation towards the densest state.  As such, elucidating the microscopic
evolution of vacancies under stress provides a dynamical perspective on the
classical close packing problem, which is a fruitful fundamental problem with
extensive applications in multiple
fields~\cite{manoharan2015colloidal,van2014understanding,
torquato2002random,weaire2008pursuit}.

The goal of this work is to explore the instability mode in the mechanical
behaviors of vacancies under compression, and to reveal the underlying
microscopic mechanism. To this end, we employ a
two-dimensional model system consisting of elastic particles of uniform size,
and resort to the combination of theoretical analysis and numerical experiment.
This model system provides the opportunity to address a host of conceptual
questions regarding the interplay of vacancy and stress, such as: How and under
which condition does a vacancy become unstable? Do vacancies of varying size and
morphology conform to a common instability mode? What is the microscopic
mechanism of vacancy instability? Illustrating these questions lays a
theoretical foundation for the strategy of vacancy based stress engineering in
extensive crystalline structures~\cite{bowick2016colloidal,porter2015granular}.
Note that while vacancies in regular crystals at non-zero temperature
equilibrium could exhibit rich physics, this work focuses on the purely
mechanical behavior of vacancies~\cite{martin1972unified, yao2014dynamics}.

In this work, we first show the concentration of stress around the vacancy by
analytical elasticity analysis. By Delaunay triangulation, we further reveal the
underlying topological defect motif of the vacancy. A vacancy consists of a pair
of dislocations. The combination of the concentrated local stress and the unique
topological structure of the vacancy constitutes the key element for the
instability of the vacancy structure.  By numerical experiments, we observe the
remarkable fission of the vacancy into a pair of dislocations under compression.
The dislocations glide along opposite directions and vanish at the
boundary. This neat process, which is triggered by the local shear stress around
the vacancy, corresponds to a highly coordinated movement of particle arrays.
Similar fission process is also observed for two- and three-point vacancies, and
in systems consisting of stiffer particles. All these observations boil down to
a common scenario of stress driven fractionalization of vacancies. This
discovery advances our understanding on the interplay of defect and stress at
the microscopic level in crystalline structures.


\section{MODEL AND METHOD}

\begin{figure}
  \centering
    \includegraphics[width=2in]{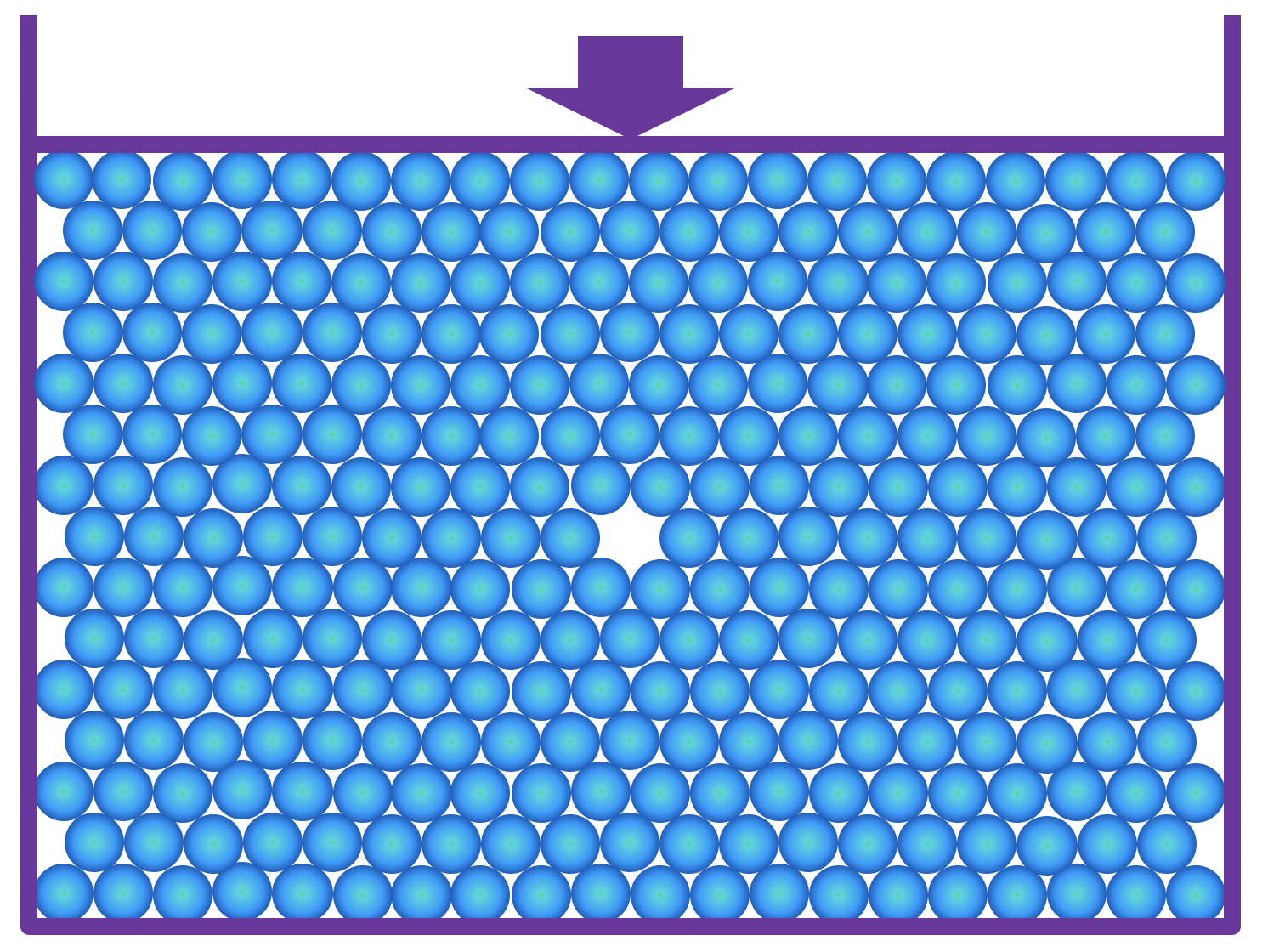}
    \par
  \protect\caption{Schematic plot of the granular crystal model system
  containing a one-point vacancy. The system is subject to compression through
  the upper movable piston. The elastic particles of uniform size form a perfect
  triangular lattice. Each particle is tangentially in contact with the
  neighboring particles or the wall in the initial stress-free configuration.
        \label{demo}}
\end{figure}

Our model consists of elastic particles with regular arrangement confined in the
two-dimensional box as shown in Fig.~\ref{demo}. The regular particle array in
the simulation box consists of $n$ layers. The number of particles in each layer
alternates between $m$ and $m-1$. In the initial state, the particle-particle
and particle-wall are in tangential contact, and the system is thus initially
stress-free.  The system is subject to
compression through a piston, which creates a uniform stress field in the
triangular lattice of particles free of vacancies.  An $n$-point vacancy is
created by removing $n$ particles in the triangular lattice.  In the background
of such a uniform, initially stress-free setting, the effect of the vacancy on
the stress pattern becomes prominent. We note that in real systems the
particles are generally not of uniform size. Size polydispersity brings
randomness,~\cite{herrmann1987violation} which could lead to nonlinear elastic
and dynamical behaviors,~\cite{goddard1990nonlinear, gilles2003low,
chong2017nonlinear} unique mechanical properties~\cite{roux2000geometric,
khalili2017numerical} and proliferation of topological defects~\cite{chaikin00a,
yao2014polydispersity}. While rich physics is underlying the mixture of
vacancies and size polydispersity, here we focus on the case of uniform particle
size to reveal the microscopic behaviors of vacancies under compression. To
highlight the effect of compression, the system is also free of friction and
gravity.

The particle-particle and particle-wall interactions are modelled by the
Hertzian potential~\cite{landau99a}.  Specifically, the particle-particle
interaction potential is
\begin{displaymath} 
    V_{{\textrm {pp}}}(r)= \left\{ \begin{array}{ll} \frac{\sqrt{2R}}{5D}(2R-r)^{\frac{5}{2}} &
    r\leq 2R \\ 
    0 &  r > 2R,
    \end{array} \right.  
\end{displaymath} 
where $r$ the distance between the centers of the particles, and $R$ is the
radius of the particle. $D = 3(1-\nu^2)/(2E)$, where $E$ is
the Young's modulus, and $\nu$ is the Poisson ratio. The interaction
potential between the particle and the rigid wall is 
\begin{displaymath} 
    V_{{\textrm {pw}}}(r)= \left\{ \begin{array}{ll} \frac{2\sqrt{R}}{5D'}(R-r)^{\frac{5}{2}} &
    r\leq R \\ 
    0 &  r > R,
        \end{array} \right.  
\end{displaymath} 
where $r$ the distance from the center of the particle to the wall, and
$D'=D/2$.  The Hertzian model has been used to investigate phase transitions in
soft particle arrays on interfaces~\cite{miller2011two} and to reveal the
screening effect of topological defect in polydispersity
phenomena~\cite{yao2014polydispersity}.

We employ continuum elasticity theory to analyze the linear mechanical response
of the system to small compression, and resort to numerical approach based on
the steepest descend method to investigate the nonlinear microscopic behaviors
of the particle array under large compression. In our numerical experiment,
the piston is pushed downward by a small displacement at each time, and the
system is then relaxed by updating the particle configuration towards the lowest
energy state. As such, the system evolves along a quasi-static process. The
variation of the height of the system is controlled within $1\%$ in each push of
the piston. We employ the steepest descend method to determine the lowest
energy particle configuration under the given compression; the particle
positions are updated collectively and individually in sequence to reduce the
energy of the system.  The typical step size is $s=10^{-3} R$.

As the termination condition, the magnitude of the total force on each particle
in the final state is required to be sufficiently small. Specifically, upon each
small push of the piston, we track the variation of the maximum magnitude of
the total force $f_{{\textrm {max}}}$ on the particles in the relaxation
process.  Typically, the evolution of particle configuration is terminated when
$f_{{\textrm {max}}}$ reduces to about $0.1\%$ of its original value. Simulation
tests show that this level of precision is sufficient for our problem.


\section{RESULTS AND DISCUSSION}

\subsection{Stress analysis of vacancy structure}

We first analyze the mechanical response of the particle array to small
compression by continuum elasticity theory. The particulate system is
approximated by a two-dimensional isotropic continuum elastic medium of Young's
modulus $E$, Poisson's ratio $\nu$, and shear modulus $\mu=E/(2(1+\nu))$. The
in-plane stress distribution is governed by the stress balance
equation~\cite{landau99a}:
\begin{eqnarray}
\partial_i \sigma_{ij}=0,
  \end{eqnarray}
where $i, j = x, y$. The system is subject to the compression force $P$ per unit
length on the piston and the shear-free boundary conditions on the three walls.
We thus have $\sigma_{xx}=-\nu P$, $\sigma_{yy}=- P$, and
$\sigma_{xy}=\sigma_{yx}=0$. The corresponding strain field is:
$u_{xx}=u_{xy}=u_{yx} =0$, $u_{yy}=-P(1-\nu^2)/E$. Here, the Poisson's ratio
$\nu$ enters the solution through the boundary condition. In general, the stress
distribution in an elastic plate subject to given forces at its edges is
independent of the elastic constants of the material~\cite{landau99a}.

\begin{figure*}[th]
    \includegraphics[width=7in]{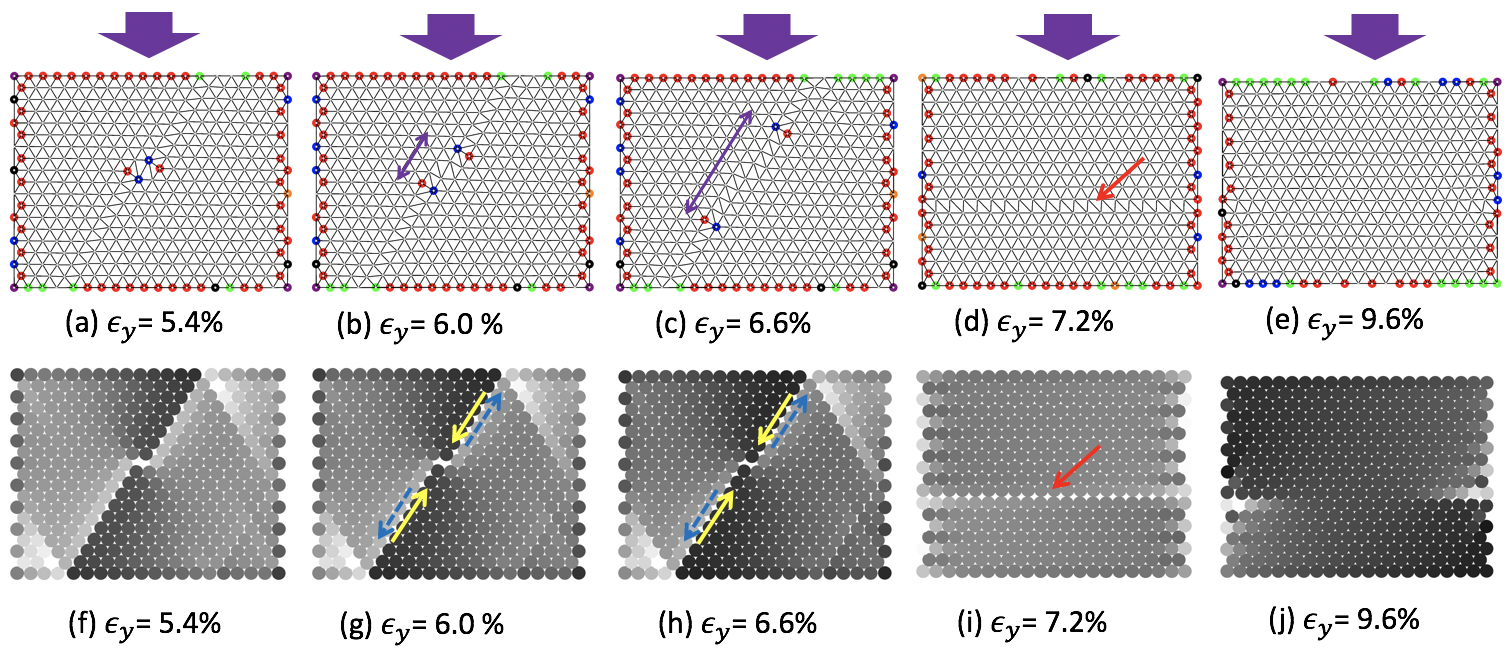}
    \par
  \protect\caption{Fractionalization process of a one-point vacancy into a pair of
  dislocations under external strain $\epsilon_y$. A dislocation is a dipole of five- and
  seven-fold disclinations. The Delaunay triangulation and elastic energy
  distribution are presented in the upper and lower figures, respectively. The red and blue
  dots represent five- and seven-fold disclinations. The green dots at the
  boundary represent four-fold disclinations. 
  Brighter particles have larger elastic energy. The solid (yellow) and dashed
  (blue) arrows in (g) and (h) indicate the collective movement of the particle
  arrays underlying the fractionalization process. The red arrow in (d) and (i)
  indicate the location of the square lattice belt that is geometrically
  incompatible with the triangular lattice. 
\label{20_19}}
\end{figure*}

How will a vacancy influence the uniform stress field in the particle array
under compression? To address this question, we consider the general case that
the two-dimensional continuum elastic medium with a circular hole of radius $a$
is horizontally compressed by the force $P_1$ per unit length, and vertically
compressed by the force $P_2$ per unit length. By linear superposition of these
two forces, the equilibrium stress field can
be derived from the stress balance equation~\cite{timoshenko1951theory,
landau99a}. The expressions
for the stress field are presented below:
\begin{eqnarray} \sigma_{rr}(r,\theta) =
    -\frac{1}{2}(P_1+P_2)(1-\frac{a^2}{r^2}) 
    -\frac{1}{2}(P_1-P_2) \nonumber \\  \times \left(1+\frac{3a^4}{r^4}
    -\frac{4a^2}{r^2}\right) \cos2\theta ,\label{sigma1}\end{eqnarray}
\begin{eqnarray} \sigma_{\theta\theta}(r,\theta) =
    -\frac{1}{2}(P_1+P_2)(1+\frac{a^2}{r^2}) 
    +\frac{1}{2}(P_1-P_2) \nonumber \\  \times \left(1+\frac{3a^4}{r^4}\right)
    \cos2\theta ,\label{sigma2}\end{eqnarray}
\begin{eqnarray} \sigma_{r\theta}(r,\theta) =
    \frac{1}{2}(P_1-P_2) \left(1-\frac{3a^4}{r^4} + \frac{2a^2}{r^2} \right)
    \sin2\theta .\label{sigma3}\end{eqnarray}
where $(r, \theta)$ are the polar coordinates with the origin at the center of
the hole structure.  Now, we apply Eqs.(\ref{sigma1})-(\ref{sigma3}) to our
system by setting $P_1=\nu P$ and $P_2= P$. The nonvanishing
$\sigma_{r\theta}$-component indicates that the hole structure is under a
shear stress field. The shear stress could lead to the glide of dislocation, and
it is related to the fractionalization of the vacancy, which will be discussed
later. Furthermore, analysis of Eq.(\ref{sigma2}) shows that the azimuthal
stress $\sigma_{\theta\theta}$ at $\theta=0$ and $\pi$ around the vacancy is
amplified by a factor of $3-\nu$ in comparison with the vacancy free case. The
concentration of stress may lead to morphological instability of the vacancy.

To characterize the discrete structure of the vacancy that is absent in the
preceding elasticity analysis, we perform a Delaunay triangulation of the
particles, and reveal a pair of dislocations as the underlying topological
structure of a compressed vacancy as shown in Fig.~\ref{20_19}(a). A dislocation
consists of a pair of five- and seven-fold disclinations, which are indicated by
the red and blue dots in Fig.~\ref{20_19}. As the elementary topological defect
in two-dimensional triangular lattice, an $n$-fold disclination refers to a
vertex whose coordination number $n$ deviates from six, and it carries a
topological charge $q=(6-n)\pi/3$.  Elasticity theory shows attraction between
oppositely charged disclinations and repulsion between like-charge
disclinations, which is in analogy with electric
charges~\cite{nelson2002defects}. A dislocation is thus analogous to an
electric dipole. The pair of dislocations in Fig.~\ref{20_19}(a) are associated
with two Burgers vectors of opposite signs; the Burgers vectors are
perpendicular to the line connecting the five- and seven-fold
disclinations~\cite{nelson2002defects}. Note that in a perfect triangular
lattice, a vacancy is represented by a ring of three
dislocations~\cite{yao2014dynamics}. A slight compression leads to the
annihilation of a dislocation; the $C_6$ symmetry is broken, but the total
topological charge is invariant in this
process~\cite{chaikin00a,yao2014dynamics}. The revealed underlying topological defect
structure of the vacancy provides the conceptual foundation for the stability
analysis of the vacancy under compression.  It is natural to ask if the
dislocation pair associated with the vacancy will be separated under the
external compression.

To address this question, we shall analyze the Peach-Koehler force acting on a
dislocation by the external stress field $\sigma^{{\textrm
(e)}}_{ij}$~\cite{peach1950forces}:
\begin{eqnarray} f_i = -\epsilon_{il}
    \sigma^{{\textrm (e)}}_{lm} b_m, \label{fi} 
\end{eqnarray} 
where $\epsilon_{ij}$ is the anti-symmetric tensor, and $\vec{b}$ is the Burgers
vector. Note that the assumption underlying Eq.(\ref{fi}) is that the external
stress field is unaffected by the presence of the dislocation, which resembles
the case of a point charge in an external electric field. The parallel component
of the Peach-Koehler force along the line of the Burgers vector drives the glide
motion of the dislocation~\cite{landau99a,chaikin00a, teodosiu2013elastic}.
Dislocation glide occurs at a much lower stress than that for the perpendicular
motion of the dislocation. The latter kind of motion is known as dislocation
climb, and it involves the addition or reduction of materials~\cite{chaikin00a,
teodosiu2013elastic}. As such, glide is much easier than climb for a dislocation
under stress. For convenience, we set the direction of the Burgers vector
$\vec{b}$ as the $x'$-axis, and introduce the $y'$-axis by counterclockwise
rotation of the $x'$-axis by $\pi/2$. For example, in the case of
Fig.~\ref{20_19}(a), the Burgers vector $\vec{b}$ of the upper dislocation,
which is perpendicular to the line connecting the five- and seven-fold
disclinations, makes an angle of $\pi/3$ with respect to the horizontal line
(the $x$-axis).

We analyze Eq.(\ref{fi}) in the $(x', y')$ coordinates. Since $m=x'$ and
$\epsilon_{ij}$ is the anti-symmetric tensor, it is required that
$\sigma^{{\textrm (e)}}_{y'x'} \neq 0$ to ensure that $f_{x'}$, the component of the
Peach-Koehler force in the $x'$-axis, is nonzero. According to the
previously solved stress field $\sigma_{ij}$ ($i,j=x,y$) under the compression
$P$, we derive for the stress field in the $(x', y')$ coordinates:
$\sigma_{x'x'}=-(3+\nu)P/4, \sigma_{y'y'}=-(3\nu+1)P/4,
\sigma_{x'y'}=\sigma_{y'x'}=\sqrt{3}(\nu-1)P/4$. We see that the external
compression produces the nonzero component $\sigma_{y'x'}$. In combination with
the analysis of Eq.(\ref{fi}), we conclude that the compression $P$ provides the
driving force for the glide of the pair of dislocations in Fig.~\ref{20_19}(a)
along opposite directions; the Peach-Koehler forces on the pair of dislocations
have opposite signs. As such, a vacancy under compression that is composed of a
dislocation pair carries the element of self-destruction upon its creation.
We resort to numerical approach to determine the critical strain at which the
fission of the vacancy occurs.  Note that the presence of friction and particle
bonding may suppress the occurrence of fractionalization.

\begin{figure}[t]
    \includegraphics[width=3.5in]{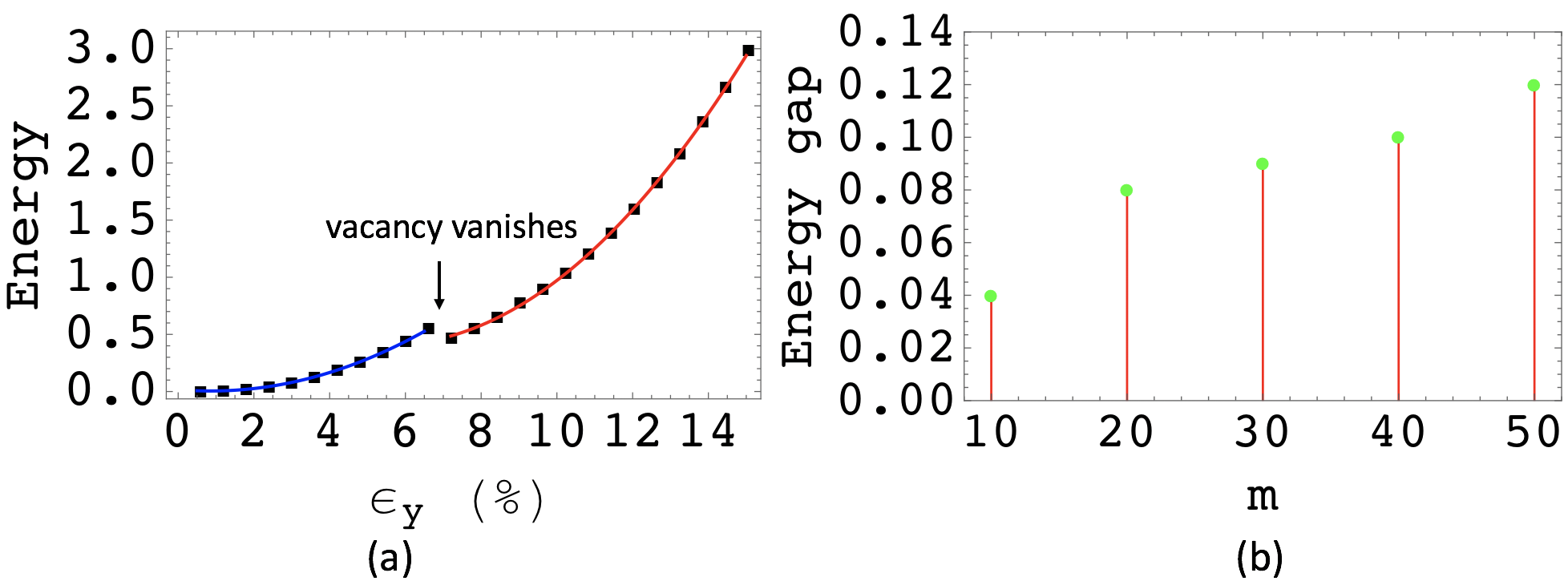}
    \par
    \protect\caption{Variation of the strain energy with
    the increase of the external strain $\epsilon_y$. (a) Plot of energy versus
    $\epsilon_y$ for the system in Fig.2. The energy gap corresponds to the
    disappearance of the vacancy. The data are well
    fitted by two quadratic curves. The coefficient of the quadratic term of the
    red fitting curve, which corresponds to the mechanical rigidity of the
    system, increases by as large as $70\%$ in comparison with that of the blue
    fitting curve. (b) Plot of the magnitude of the energy gap versus
    the size of the system. $n=m-1$.  
  \label{energy_strain}}
\end{figure}

\subsection{Fractionalization of vacancies}

In Fig.~\ref{20_19}, we show the morphological transformation of the vacancy
with the increase of the external strain $\epsilon_y$. A vacancy under
compression is represented by a pair of dislocations; each dislocation consists
of a five- and seven-fold disclinations, as indicated by the red and blue dots
in Fig.~\ref{20_19}(a). Figure~\ref{20_19}(b) shows that
fractionalization of the vacancy occurs when $\epsilon_y$ reaches about $6\%$.
Under stronger compression, the dislocations continuously glide and ultimately
vanish at the boundary in response to a small change of the external strain
$\epsilon_y$. Notably, this instability mode persists as the system size and the
vacancy location are changed.

In Fig.~\ref{20_19}(d), we notice the appearance of a square lattice belt
indicated by the red arrow.  This geometrically incompatible structure embedded
in the triangular lattice is eliminated by a further increase of the strain,
leaving out a perfect triangular lattice as shown in Fig.~\ref{20_19}(e). We
also notice that the number of layers decreases from 19 to 18 from
Fig.~\ref{20_19}(c) to Fig.~\ref{20_19}(d). This layering transition, which is
triggered mechanically, occurs with the vanishing of the dislocations at the
boundary; the number of layers is invariant as vacancy fractionalization occurs
from Fig.~\ref{20_19}(a) to Fig.~\ref{20_19}(c).  Layering transition driven by
thermal fluctuations is commonly seen in confined solids of hard spheres and
colloidal particles where there are no vacancies~\cite{pieranski1980hard,
schmidt1996freezing, chaudhuri2004constrained, ricci2006lack, fortini2006phase,
ricci2007ordering}. As such, the presence of vacancies is not a necessary
condition for the occurrence of layering transition. Both vacancy
fractionalization and layering transition require sufficiently high strain. In
connection with real systems, high strain may be realized in a series of
deformable mesoparticle systems like ultrasoft colloids and hydrogel beads; both
soft interaction and particle size can be engineered in these soft particle
systems~\cite{Likos2006, Miller2011, girardo2018standardized}.

To rationalize the numerically observed fractionalization of the vacancy, we
present analytical analysis based on the concepts of resolved shear stress and
passing stress~\cite{veyssiere2007equilibrium, teodosiu2013elastic}.  Resolved
shear stress refers to the component of the stress in the glide direction of a
dislocation. The resolved shear stress to split a dislocation dipole of
separation $h$ is referred to the passing stress. By linear elasticity theory,
the passing stress is $\tau_P=\mu b/(8\pi(1-\nu)h)$, where $\mu$ and $\nu$ are
the shear modulus and Poisson ratio, and
$b=|\vec{b}|$~\cite{veyssiere2007equilibrium}. From the previously solved stress
field $\sigma_{ij}$ ($i,j=x,y$), we compute for the magnitude of the resolved
shear stress as $|\tau|=\sqrt{3}(1-\nu)P/4$.  Inserting the critical value for
$P$ to the previously derived expression for the strain $u_{yy}$, and with $h=b$
and $\nu=1/3$ (for 2D triangular lattice composed of linear springs), we
estimate the passing strain $u_{yy,P}$ as 
\begin{eqnarray}
  u_{yy,P} = \frac{1}{4\sqrt{3}\pi}\frac{1}{1-\nu} = \frac{\sqrt{3}}{8\pi}
  \approx 6.89\%,
\end{eqnarray}
which is close to the numerically observed critical strain ($6\%$).

Here, we shall emphasize that the observed instability mode of the vacancy is
specific to two-dimensional systems. It is of interest to investigate the
transformation of vacancies under compression in three-dimensional systems,
which is beyond the scope of this work. The stress driven fractionalization of
the vacancy is fundamentally different from the thermally driven diffusive
motion of the vacancy as a whole in a two-dimensional Lennard-Jones crystal; the
integrity of the vacancy is well preserved and no fission occurs in the latter
case~\cite{yao2014dynamics}. Curvature driven fractionalization of interstitial
in two-dimensional crystal lattice has also been
reported~\cite{bowick2007interstitial,irvine2012fractionalization}. Here,
fractionalization occurs on vacancies and is driven by external stress, which
enlarges the space of manipulation in stress and defect
engineering~\cite{bowick2016colloidal}.

From the distribution of the strain energy in
Fig.~\ref{20_19}(f)-\ref{20_19}(h), where the brighter particles have larger
elastic energy, we see that the fractionalization process of the vacancy is
realized by the highly coordinated movement of particle arrays as indicated by
the solid (yellow) and dashed (blue) arrows. The vacancy driven heterogeneous
stress distribution also demonstrates the intrinsic connection between defects
and the formation of force chains in granular matter~\cite{liu1998nonlinear}. In
the rearrangement of the particles around the vacancy, the total energy of the
system experiences an abrupt reduction, as shown in Fig.~\ref{energy_strain}(a).
Figure~\ref{energy_strain}(a) also shows a significant stiffening of the system
with the disappearance of a single one-point vacancy.  The coefficient of the
quadratic term of the red fitting curve, which corresponds to the mechanical
rigidity of the system, increases by as large as $70\%$ in comparison with that
of the blue fitting curve. In Fig.~\ref{energy_strain}(b), we show that the
magnitude of the energy gap increases with the size of the system.

\begin{figure}[t]  
\centering 
\includegraphics[width=3.6in]{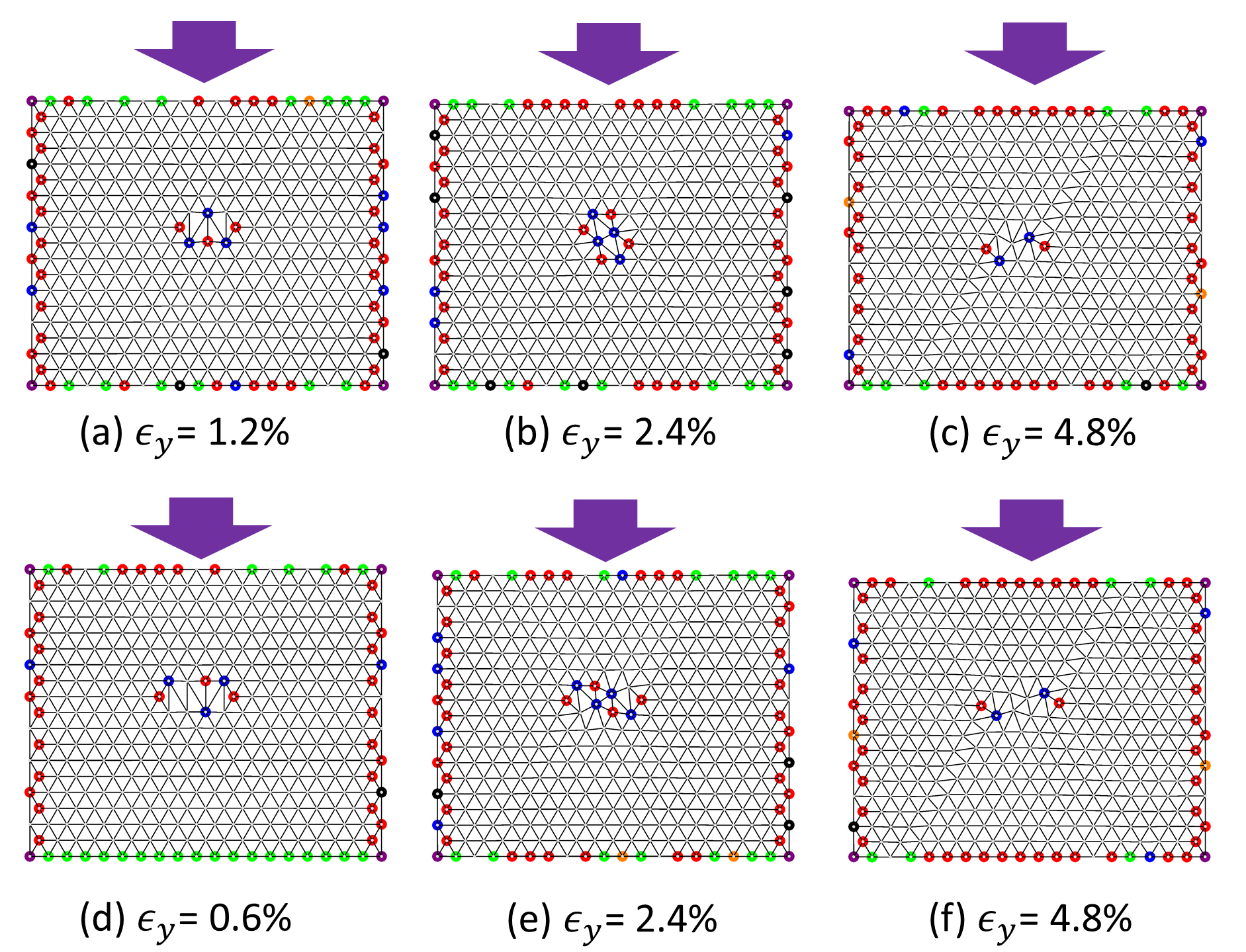}
\caption{Fractionalization of the two- and three-point vacancies into a single
  pair of dislocations under compression. (a)-(c) Morphological transformation
  of the two-point vacancy under varying external strain $\epsilon_y$. (d)-(f)
  The case of the three-point vacancy.  The red and blue dots represent five-
  and seven-fold disclinations. The green dots at the
  boundary represent four-fold disclinations.  }
\label{two_three_pt_vac_horizon}
\end{figure}

\begin{figure*}[th]  
\centering 
\includegraphics[width=7in]{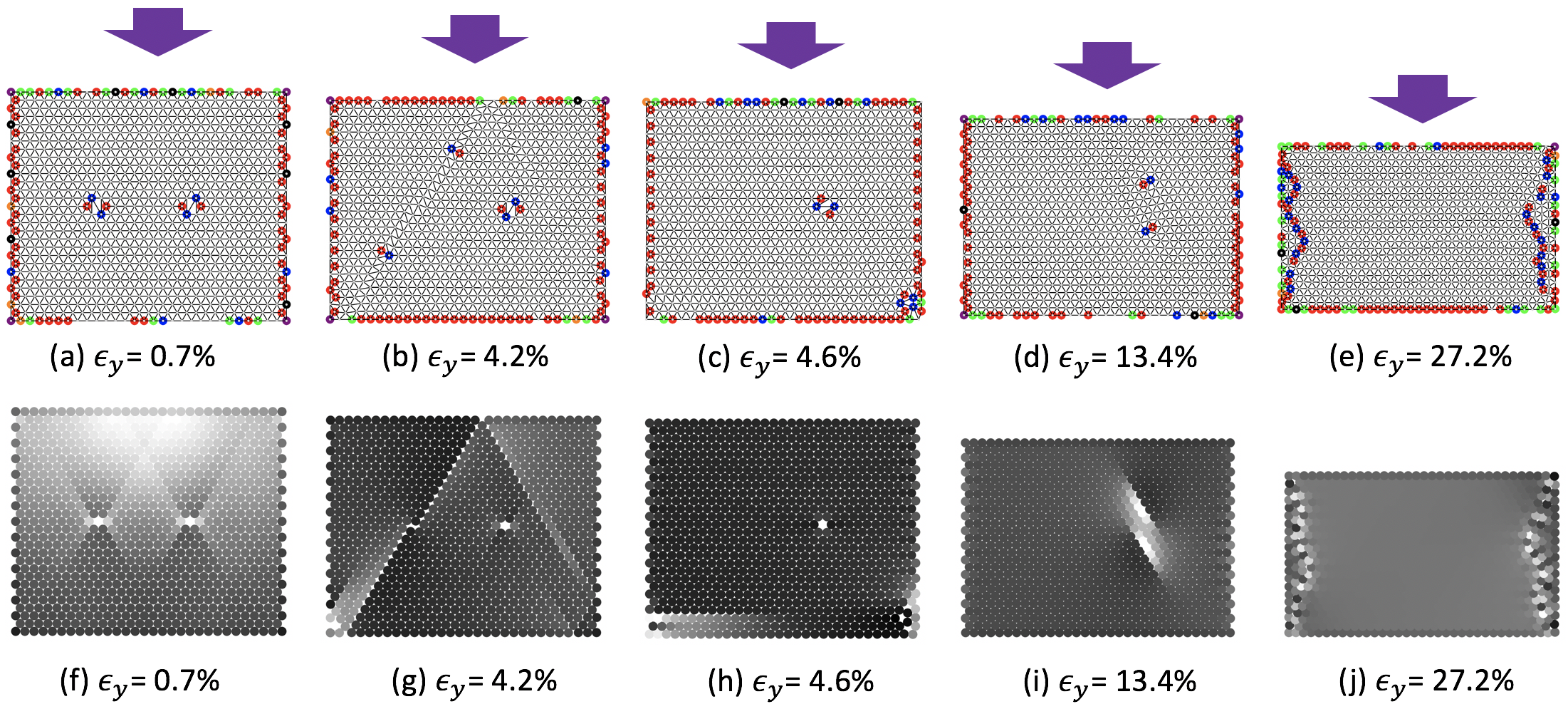}
\caption{Consecutive fractionalization of two one-point vacancies under
  compression. The Delaunay triangulation and elastic energy distribution are
  presented in the upper and lower figures, respectively. The red and blue dots
  represent five- and seven-fold disclinations. The green dots at the
  boundary represent four-fold disclinations. Brighter particles have larger
  elastic energy.  }
\label{two_vac}
\end{figure*}

\begin{figure}[h]
  \centering
    \includegraphics[width=2.5in]{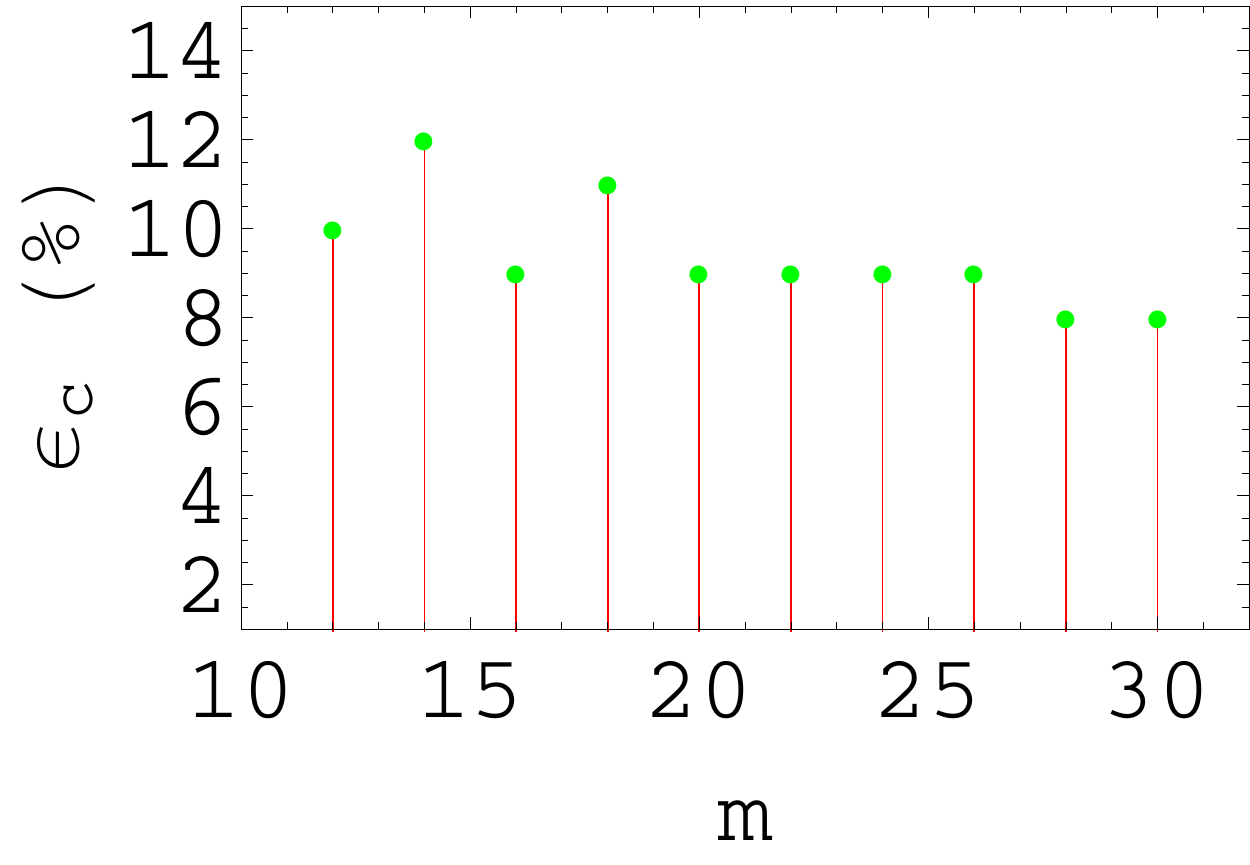}
    \par
	\protect\caption{Plot of the critical strain $\epsilon_c$ versus the size
        of the system. A one-point vacancy is introduced at the center of the
particle array. $n=m-1$. 
    \label{ec_vs_n}}
\end{figure}

We proceed to discuss the microscopic evolution of two- and three-point
vacancies under compression. The main results are presented in
Fig.~\ref{two_three_pt_vac_horizon}. Like the case of one-point vacancy in
Fig.~\ref{20_19}, both two- and three-point vacancies are uniformly fissioned
to a single pair of dislocations under compression regardless of their
morphology.  Figure~\ref{two_three_pt_vac_horizon} shows that the disclinations
(the colors dots) around the vacancies experience a series of reconfiguration
and annihilation events, and converge to a single pair of dislocations prior
to fission. Simulations show that two- and three-point vacancies with varying
orientations and morphologies conform to the same fractionalization scenario as
in Fig.~\ref{two_three_pt_vac_horizon}. Notably, the critical strain for the
fractionalization of the two- and three-point vacancies is reduced to about
$4.8\%$, which is much lower than the value of $6\%$ for the one-point vacancy
case. We further investigate the microscopic behavior of multiple one-point
vacancies under compression. Figure~\ref{two_vac} shows the consecutive
fractionalizations of the vacancies into pairs of dislocations that ultimately
vanish at the boundary. The critical strain reduces to about $4\%$. We also
place the vacancy at different layers and still observe vacancy
fractionalization.  To conclude, the instability mode of vacancies of varying
size and morphology conforms to a common fractionalization scenario.

\subsection{Effects of particle stiffness, system size and simulation
parameters}

Finally, we briefly discuss the effect of particle stiffness on the instability
of vacancies, and the dependence of the critical strain on system size and
relevant simulation parameters. 

The phenomenon of vacancy fractionalization is also found in systems consisting
of stiffer particles. By increasing the power of $5/2$ in the expressions for
$V_{{\textrm {pp}}}(r)$ and $V_{{\textrm {pw}}}(r)$ to $7/2$, the critical
strain for the fractionalization of the vacancy increases accordingly from
about $6\%$ to $10\%$.  The increase of the critical strain can be attributed to
the enhanced binding energy of the dislocation pair that is proportional to the
Young's modulus~\cite{chaikin00a}. The stiffening of the particles leads to the
enhanced Young's modulus and thus the increase of the binding energy of the
dislocation pair. Stronger external stress is therefore required to pull a
dislocation pair apart.

Regarding the dependence of the critical strain on the number
of particles and relevant simulation parameters, the results are summarized in
Fig.~\ref{ec_vs_n} and Table~\ref{table:ec}, respectively. From
Fig.~\ref{ec_vs_n}, we see that the value for the critical strain weakly depends
on the number of particles. We also investigate the influence of the step size
and the termination condition on the value of the critical strain, and present
the results in Table~\ref{table:ec}. It shows that the variation of these
simulation parameters by an order of magnitude only leads to a $1\%-2\%$
variation of the value for the critical strain.

\begin{center}
    \begin{table}[h]
\caption{The value for the critical strain $\epsilon_c$ obtained at varying
      termination condition $\tau$ and step sizes $s$. $\tau = 0.001$ while
      varying $s$, and $s=0.001$ while varying $\tau$. $m=20$. $n=19$. } 
\centering
\begin{tabular}{|c|c||c|c|}
  \hline
  $\tau$ & $\epsilon_c$ & $s$ & $\epsilon_c$ \\
  \hline
  0.001 & 7\%  & 0.0005 & 8\% \\
  \hline
  0.005 & 8\%  & 0.001 & 9\% \\
  \hline
   0.01 & 9\%  & 0.005 & 8\% \\
  \hline
   0.05 & 9\%  & 0.001 & 9\% \\
  \hline
    0.1 & 10\% & 0.005 & 9\% \\
  \hline
\end{tabular}
\label{table:ec}
\end{table}
\end{center}

\section{Conclusions}

In summary, we revealed a common instability mode of vacancies with varying
size and morphology under compression. Triggered by the local shear
stress, the vacancy is fissioned into a pair of dislocations that ultimately
glide to the boundary. The remarkable fractionalization of vacancies creates
rich modes of interaction between vacancies and other defects,
and provides a new dimension for mechanical engineering of defects in extensive
crystalline materials. This discovery may also be exploited to explore the
classical close packing problem from the dynamical perspective by using the
softness of particles to eliminate vacancies and achieve higher particle
density.

\section{Acknowledgements}
This work was supported by NSFC Grants No. 16Z103010253, the SJTU startup fund
under Grant No. WF220441904. The author thanks the support from the Student
Innovation Center at Shanghai Jiao Tong University.


\end{document}